\documentclass[aip,pra,twocolumn,longbibliography]{revtex4}
\usepackage{blindtext}
\usepackage[usenames,dvipsnames]{color}
\usepackage{graphicx}
\usepackage{dcolumn}
\usepackage{bm}
\usepackage{amsmath}
\usepackage{amsfonts}
\usepackage{psfrag}
\usepackage{mathtools}
\usepackage{accents}

\usepackage[mathcal]{euscript}

\begin{document}

\title{Quantum theory of longitudinal-transverse polaritons in nonlocal thin films}

\author{Christopher R. Gubbin}
\author{Simone De Liberato}
%\email[Corresponding author: ]{s.de-liberato@soton.ac.uk}
\affiliation{School of Physics and Astronomy, University of Southampton, Southampton, SO17 1BJ, United Kingdom}

\begin{abstract}
When mid-infrared light interacts with nanoscale polar dielectric structures optical phonon propagation cannot be ignored, leading to a rich nonlocal phenomenology which we have only recently started to uncover. In properly crafted nanodevices this includes the creation of polaritonic excitations with hybrid longitudinal-transverse nature, which are predicted to allow energy funnelling from longitudinal electrical currents to far-field transverse radiation. In this work we study the physics of these longitudinal-transverse polaritons in a dielectric nanolayer in which the nonlocality strongly couples epsilon-near-zero modes to longitudinal phonons. After having calculated the system's spectrum solving Maxwell's equations, we develop an analytical polaritonic theory able to transparently quantify the nonlocality-mediated coupling as a function of the system parameters.
Such a theory provides a powerful tool for the study of longitudinal-transverse polariton interactions and we use it to determine the conditions required for the hybrid modes to appear. 
\end{abstract}
\maketitle

% General nanophotonic intro
Photonic energy can be confined to deep sub-diffraction lengthscales  by hybridisation of light with optically active transitions \cite{Ballarini2019}, thus storing part of the electromagnetic energy in the charges' kinetic energy \cite{Khurgin2017}. In the mid-infrared region this can be achieved by coupling light with the transverse optical phonons  of a polar nanostructure, yielding hybrid light-matter excitations termed surface phonon polaritons (SPhPs) \cite{Greffet2002,Hillenbrand2002,Caldwell2015a}. These modes are highly tuneable \cite{Sumikura2019, Spann2016, Ellis2016,Gubbin2017,Dubrovkin2020} and have broad applications in nonlinear optics \cite{Gubbin2017b, Razdolski2018,Kitade2021}, near-field imaging \cite{Taubner2006, Kiessling2019}, design of mid-infrared emitters \cite{Wang2017} and fabrication of nanophotonic circuitry \cite{Li2018,Chaudhary2019, Li2016}.\\
% Strong coupling between different modes (ENZ,NL)
Due to their narrow linewidths and large field confinement SPhPs can be strongly coupled to a variety of other resonances present in nanostructured devices. Polaritonic excitations resulting from such strong coupling are hybrid quasiparticles whose unique properties can be understood as admixtures of those of their bare components. 
Resonances which have been experimentally demonstrated to be strongly coupled with SPhPs include localised phonon polaritons \cite{Gubbin2016},  epsilon-near-zero (ENZ) modes \cite{Passler2018}, weak phonon excitations \cite{Gubbin2019}, and more recently vibrational transitions in organic molecules \cite{Bylinkin2021}. \\
Of particular relevance for the present work, in Ref.~\cite{Gubbin2019} SPhPs localised in silicon carbide (SiC) nanopillars were demonstrated to be strongly coupled to longitudinal optical (LO) phonons whose dispersion is folded along the c-axis of the 4H-SiC polytype. The resulting excitations were named longitudinal-transverse polaritons (LTP) because they have the unusual feature of possessing both a longitudinal nature from their LO component, and a transverse one from their electromagnetic SPhP component. These recently-observed LTP quasiparticles are an exciting system for applications in mid-infrared optoelectronics, with a transverse photonic component able to radiate to the far-field and a longitudinal matter component able to interact with electrical currents. They could enable a novel generation of efficient and low-cost mid-infrared optoelectronic devices, not requiring the usual quantum cascade structures to convert electrical currents into mid-infrared radiation \cite{Ohtani2019}. Note that longitudinal-transverse strong coupling has recently also been observed in fluids \cite{Kryuchkov2019}.\\
\begin{figure}[t!]
\begin{center}
	\includegraphics[width=0.5\textwidth]{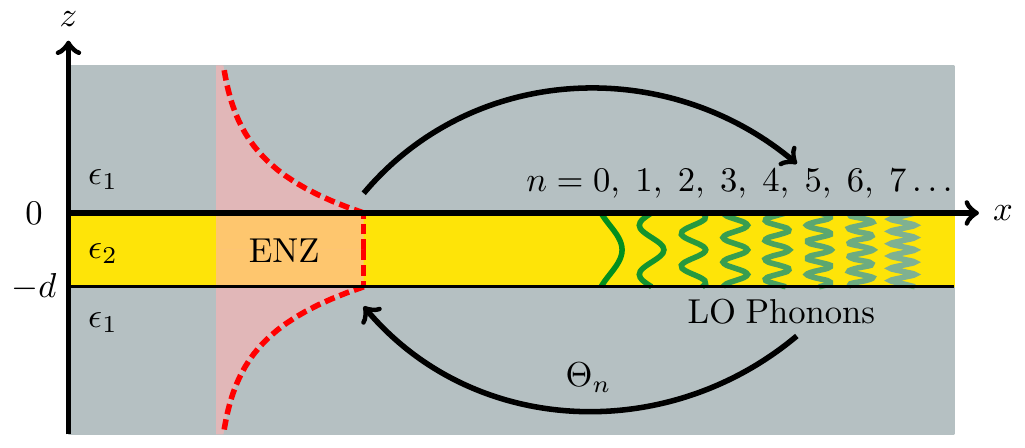}
	\caption{\label{fig:Fig0} The waveguide under study in this work. Illustrated are the out of plane electric field $\mathrm{E}_z$ for the ENZ mode (red) and localised phonons (green). The coupling frequency $\Theta_n$ couples the ENZ to the $n$th longitudinal phonon mode.}
\end{center}
\end{figure}
As we showed in a series of recent papers \cite{Gubbin2020,Gubbin2020b,Gubbin2020c} the coupling between longitudinal and transverse degrees of freedom can be more generally understood as a feature of nanoscopic polar devices where optical phonon propagation cannot be neglected. The usual local dielectric approximation then fails and the resulting nonlocal phenomenology driving the longitudinal-transverse coupling has important consequences on the performances of polar nanodevices, analogous to nonlocal effects in localised plasmons \cite{Ciraci2012,Ciraci2013} and plasmon polaritons \cite{Rajabali2021}. \\
Polaritonic systems are usually studied using second quantized approaches \cite{Hopfield1958,Alpeggiani2014,Gubbin2016b}, which were very successful, e.g., in the study of polariton-polariton \cite{Carusotto2013,Gubbin2017b} and electron-polariton \cite{Tan2020,Efimkin2021} interactions. These are important for LTP systems, the former would allow for the description of exotic phonon polariton condensates, while the latter would be especially important for the development of phonon-polariton optoelectronics, allowing for modelling of both the excitation of LTPs from electrical currents and their outcoupling to the far-field. Extensions of the Hopfield polaritonic formalism to general nanophotonic devices, and even more to LTP, becomes nevertheless problematic when the coupled resonances are morphology-dependent. The bare uncoupled modes can then be difficult to identify or altogether ill-defined. In these cases only an holistic solution of the Maxwell equations coupled to spatially varying dielectrics could seem to provide a predictive description of the hybrid resonances, thus foregoing the explanatory and operational power of polaritonic formalism.\\
In this work we achieve two objectives which we consider timely and relevant to advance our understanding and modelling capabilities of solid-state mid-infrared nanophotonic devices. (I) Using the approach we developed in Ref.~\cite{Gubbin2020}, that is solving the full system of Maxwell equations in the nonlocal dielectric and taking mechanical boundary conditions into account, we study the nonlocal physics of a thin nanolayer, in which LTP are formed by the strong coupling of the layer's epsilon-near-zero resonance  to the layer's discrete LO phonons. 
(II) We develop an analytical second-quantized polaritonic formalism, allowing us to describe the system in terms of coupled harmonic oscillators. The solution of the nonlocal problem then requires only the bare frequencies of the modes in the range of interest and their couplings. 
This is a powerful theoretical tool for mid-infrared nanophotonics, partially alleviating the complexity of the nonlocal electromagnetic theory \cite{Gubbin2020c}. It allows us for example to identify the the geometric and material parameters required to achieve strong coupling between longitudinal and transverse modes.
Moreover, the polaritonic solution describes the system in terms of hybrid quasiparticles whose interactions can be studied with tested tools both below and above the lasing threshold \cite{Carusotto2013}. It is thus a key tool for future studies of the LTP-mediated coupling between electrical currents and far-field mid-infrared electromagnetic radiation.
%%% End of introduction 
 
\begin{figure}[t!]
\begin{center}
	\includegraphics[width=0.35\textwidth]{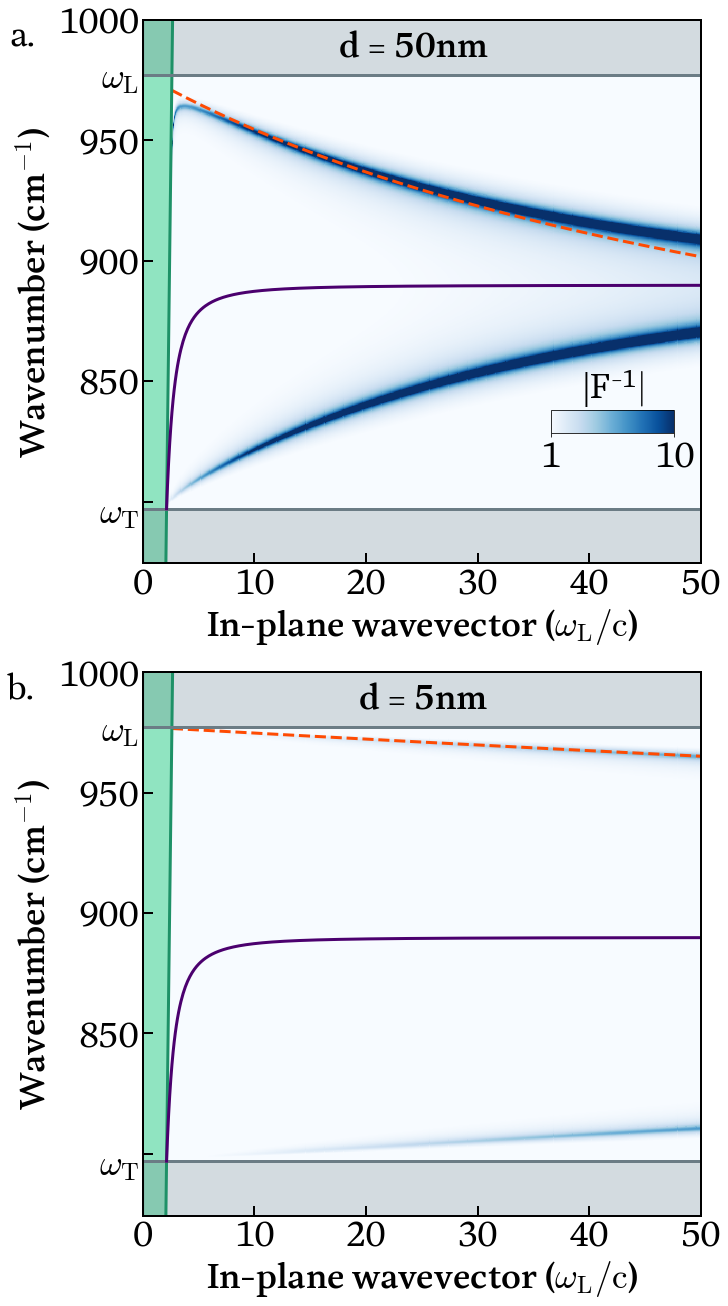}
	\caption{\label{fig:Fig1} Modal dispersion for a 50nm (a) and 5nm (b) 3C-SiC waveguide between high index ($n=2.6$) cladding layers. The white region illustrates the spectral extent of the Reststrahlen between $\omega_{\mathrm{T}}$ and $\omega_{\mathrm{L}}$, where SPhPs are supported. The leaky region within the light-line is shaded in green. The analytical ENZ dispersion from Eq.~\ref{eq:enzdisp} is shown by the dashed red line, and the bilayer SPhP dispersion by the solid purple curve.}
\end{center}
\end{figure}

%%% System description
We consider the trilayer waveguide shown in Fig.~\ref{fig:Fig0}. This consists of a polar film of thickness $d$ and relative permittivity $\epsilon_2$ sandwiched between identical semi-infinite regions with identical relative permittivity $\epsilon_{1}$. The waveguide dispersion in the local response approximation can be derived considering TM polarised electromagnetic fields and applying electromagnetic boundary conditions on the components of the magnetic and electric fields parallel to the surface $\mathrm{H}_{\parallel}$ and $\mathrm{E}_{\parallel}$ \cite{Maierbook}, leading to eigenequation 
\begin{equation}
	\mathrm{F} = 2 - \tanh\left(\alpha_2 d\right) \left[ \frac{\epsilon_2 \alpha_1}{\epsilon_1 \alpha_2} + \frac{\epsilon_1 \alpha_2}{\epsilon_2 \alpha_1}\right] = 0, \label{eq:locdisp}
\end{equation}
where $\alpha_i^2 = k_{\parallel}^2 - \epsilon_i \omega^2 / c^2$ is the out-of-plane wavevector in the $i$th layer. 
We consider region $1$ as a positive lossless dielectric, while region $2$ is a dissipative polar dielectric characterised by a Lorentzian dielectric function 
\begin{equation}
\epsilon_2\left(\omega\right) = \epsilon_{\infty} \left[\frac{\omega_\mathrm{L}^2 - \omega \left(\omega + i \gamma\right)}{\omega_{\mathrm{T}}^2 - \omega \left(\omega + i \gamma\right)}\right],
\end{equation}
where $\omega_{\mathrm{L}} \; (\omega_{\mathrm{T}})$ is the longitudinal (transverse) bulk optical phonon frequency, $\epsilon_{\infty}$ is the high-frequency dielectric constant, and $\gamma$ is the phonon loss rate, which for the sake of simplicity we consider frequency and polarization independent. \\
A thick polar film acts as a semi-infinite medium, supporting isolated SPhP modes on each interface. These modes exist within the Reststrahlen region $\omega_{\mathrm{T}} < \omega < \omega_{\mathrm{L}}$ where $\epsilon_2(\omega) < 0$, illustrated by the non-shaded spectral region in Fig.\ref{fig:Fig1}a, b. Their dispersion is shown by the purple-solid curve. When film thickness $d$ is less than the skin depth the SPhPs on each film interface hybridise into symmetric and antisymmetric modes. This is shown in Fig.~\ref{fig:Fig1}a,b, where we plot dispersion $\lvert\mathrm{F}^{-1}\rvert$ for 50nm and 5nm 3C-SiC films respectively, using parameters $\omega_{\mathrm{T}} = 797.0\mathrm{cm}^{-1}$, $\omega_{\mathrm{L}} = 977.3 \mathrm{cm}^{-1}$, $\epsilon_{\infty} = 6.49$ and $\gamma = 2\mathrm{cm}^{-1}$ \cite{Moore1995,Debernardi1999}.
The new modes, illustrated by peaks in the colormap, split around the bare SPhP frequency and as the film thins are pushed closer to asymptotic ($d\rightarrow 0$) frequencies at $ \omega_{\mathrm{L}}$ and $\omega_{\mathrm{T}}$.

\begin{figure*}[t!]
\begin{center}
	\includegraphics[width=0.8\textwidth]{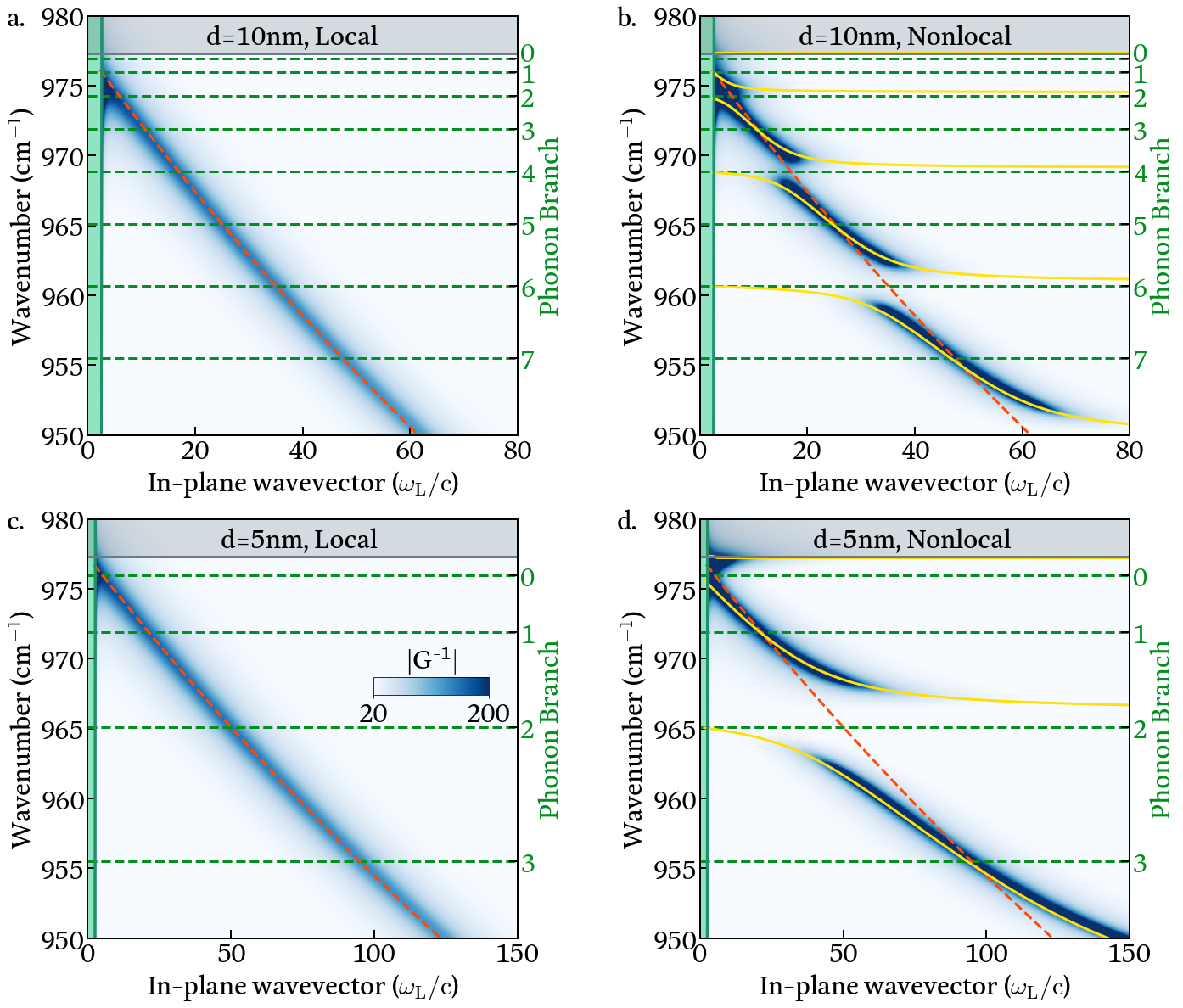}
	\caption{\label{fig:Fig2} Local (a) and nonlocal (b) dispersion relations for a 10nm SiC film. Localised phonon frequencies $\omega_{n, k_{\parallel}}$ from Eq.~\ref{eq:phononfreq} are illustrated with green dashed lines, with darker colors corresponding to higher order modes. The red dashed line indicates the ENZ from Eq.~\ref{eq:enzdisp}.
	Solid yellow lines indicate calculated polariton frequencies. Lower panels (c, d) show the same for a 5nm SiC film. The colorscale is uniform across all panels.}
\end{center}
\end{figure*}

In the limit $\alpha_2 d \ll 1$, focusing on the symmetric blue-shifted mode, we can re-write Eq.~\ref{eq:locdisp} as
\begin{equation}
\label{eq:ENZ}
	 \frac{\epsilon_2}{\epsilon_1} + \frac{k_{\parallel} d}{2} = 0,
\end{equation}
where we took the small argument limit $\tanh\left(\alpha_2 d\right) \approx \alpha_2 d$, assumed large in-plane wavevector $\alpha_1  \approx k_{\parallel}$, and noted that close to $\omega_{\mathrm{L}}$ we have $\epsilon_2\left(\omega\right) \approx 0$ and thus $\epsilon_2 \left(\omega\right) k_0^2 \ll k_{\parallel}^2$, and $\epsilon_2\left(\omega\right)^2 \ll \epsilon_1^2$. In this regime the excitation is typically termed an epsilon-near-zero (ENZ) mode \cite{Campione2015}. These are of great interest as the vanishing dielectric function $\epsilon_2 \to 0$ leads to a very strong enhancement in the out-of-plane field in the thin film with applications in low loss waveguiding \cite{Passler2018} and vibrational spectroscopy \cite{Folland2020}. Solving for $\omega$ we can find the ENZ mode frequency as a function of in-plane wavevector $k_{\parallel}$
\begin{equation}
	\tilde{\omega}_{\mathrm{ENZ}, k_{\parallel}} = \sqrt{\frac{\omega_{\mathrm{L}}^2 + k_{\parallel} q \omega_{\mathrm{T}}^2}{1 + k_{\parallel} q} - (\gamma / 2)^2} - \frac{i \gamma}{2}, \label{eq:enzdisp}
\end{equation}
where $q = d \epsilon_1 / 2 \epsilon_{\infty}$ and we use the tilde to indicate a complex frequency. This result is valid for a finite range of in-plane wavevectors because, as discussed above, when $k_{\parallel}$ becomes large the small argument approximation breaks down. The calculated ENZ dispersion $\omega_{\mathrm{ENZ}, k_{\parallel}}$ is shown by red-dashed lines in Fig.~\ref{fig:Fig1}a, b, where it is clear that the approximations from Eq.~\ref{eq:enzdisp} are less accurate for thicker films and larger in-plane wavevectors. \\

Nonlocal effects in polar dielectics have been studied both theoretically \cite{Gubbin2020,Gubbin2020c, Rivera2018} and experimentally \cite{Ratchford2019} in layered nanostructures called hybrid crystals. Importantly in a nonlocal model phonons do not exist solely at fixed frequencies, instead they are dispersive. Solving the ionic equation of motion (Eq.~A1 of Ref. \cite{Gubbin2020}) for longitudinally polarised modes leads to complex out-of-plane wavevectors
\begin{equation}
 \xi = \sqrt{\frac{\omega_{\mathrm{L}}^2 - \omega \left(\omega + i \gamma\right)}{\beta_{\mathrm{L}}^2} - k_{\parallel}^2}, \label{eq:LOFull}
\end{equation}
where $\omega_{\mathrm{L}}$ is the bulk LO phonon frequency, $\omega$ is the driving frequency, $\beta_{\mathrm{L}}$ is the LO phonon propagation velocity and $k_{\parallel}$ is the in-plane wavevector. In Ref.~\cite{Gubbin2020} we demonstrated that thin polar films of width $d$ clad by phonon inactive layers act as Fabry-P{\'e}rot resonators for LO phonons, supporting discrete modes whose real ($\gamma \to 0$) resonant frequencies can be written
\begin{equation}
	\omega_{n, k_{\parallel}} = \sqrt{\omega_{\mathrm{L}}^2 - \beta_{\mathrm{L}}^2 \left(\xi_n^2 + k_{\parallel}^2\right)}, \label{eq:phononfreq}
\end{equation}
where $\xi_n = (n + 1) \pi / d$ is the discrete out-of-plane phonon wavevector \cite{Gubbin2020c}, this notation has been chosen for clarity, in order that algebraic and geometric parity of the modes is the same. Note that, contrary to photonic Fabry-P{\'e}rot resonators, higher order phonon modes are red-shifted due to the negative dispersion of optical phonons at small wavevectors. We plot the spectrum $\omega_{n, k_{\parallel}}$ for 3C-SiC films in Fig.~\ref{fig:Fig2}, using horizontal dashed lines for 10nm (top panels) and 5nm (bottom panels) film thicknesses. Here we use 3C-SiC phonon velocity $\beta_{\mathrm{L}} = 15.39 \times 10^5 \mathrm{cm \; s}^{-1}$, derived from the bulk phonon dispersion \cite{Karch1994}. Note that within the interval of in-plane wavevectors considered $\beta_{\mathrm{L}}k_{\parallel}\ll \omega_{\mathrm{L}}-\omega_{\mathrm{T}}$ the phonons do not appreciably disperse, for this reason we approximate longitudinal mode frequencies by their zone-centre values $\omega_{n} \approx \omega_{n, 0}$.
In the same Figures the local ENZ dispersion from Eq.~\ref{eq:enzdisp} is shown by a red-dashed line. In thinner films out-of-plane momentum $\xi_n$ grows and phonon branches red shift away from $\omega_{\mathrm{L}}$.\\
We are interested in the nonlocal response of the ENZ excitation. We also consider longitudinal phonons in the central layer, with electric field given by
\begin{align}
	\mathbf{E}_{\mathrm{L}}^{\pm} = \frac{ e^{i k_{\parallel} x} }{\sqrt{2}}  \left[\left(\begin{array}{c}
 	1\\
 	0\\
 	- i \xi / k_{\parallel}	
 \end{array}\right)e^{\xi z}  \pm \left(\begin{array}{c}
 	1\\
 	0\\
 	i \xi / k_{\parallel}	
 \end{array}\right)e^{-\xi (z + d)}\right], \label{eq:LOField}
\end{align}
where the $\pm$ labels the parity of the phonon field. We refer to parity with respect to the out-of-plane electric field. The ENZ mode has constant electric field across the film so it's parity is even.\\
We demonstrated in Ref. \cite{Gubbin2020} that the full nonlocal dispersion can be found applying the Maxwell boundary conditions on the in-plane field components $\mathrm{H}_{\parallel}, \mathrm{E}_{\parallel}$ and a single nonlocal additional boundary condition on the out-of-plane one $\epsilon_{\infty} \mathrm{E}_{\perp}$. These conditions lead to a nonlocal dispersion relation $\mathrm{G}=0$, where
\begin{multline}
    		\mathrm{G} = \tanh\left(\frac{\alpha_2 d}{2}\right) + \frac{\alpha_1 \epsilon_2}{\alpha_2 \epsilon_1} \\
    		+ \tanh\left( \frac{\xi d}{2}\right) \left(\frac{\epsilon_2}{\epsilon_{\infty}} - 1\right) \frac{k_{\parallel}^2}{\xi \alpha_2}. \label{eq:nldisp}
\end{multline}
Note that here only phonons of odd-parity ($-$ branch of Eq.~\ref{eq:LOField}) are hybridised to the odd-parity ENZ field. We plot the inverse of the dispersion $\lvert\mathrm{G}^{-1}\rvert$ as a colormap in Fig.~\ref{fig:Fig2}.
Left panels (a,c) correspond to the local ($\beta_{\mathrm{L}}=0$) case, while right ones (b,d) show the nonlocal one. Top (a,b) and bottom (c,d) panels correspond to 10nm and 5nm SiC films respectively. Compared to the local case we see nonlocality leads to the appearance of anti-crossings between the ENZ mode and even-numbered localised phonon branches, with increasing coupling for higher $n$ phonons. These even numbered branches correspond to solutions where the out-of-plane electric field vanishes at the film edge.
\begin{figure}
\begin{center}
	\includegraphics[width=0.45\textwidth]{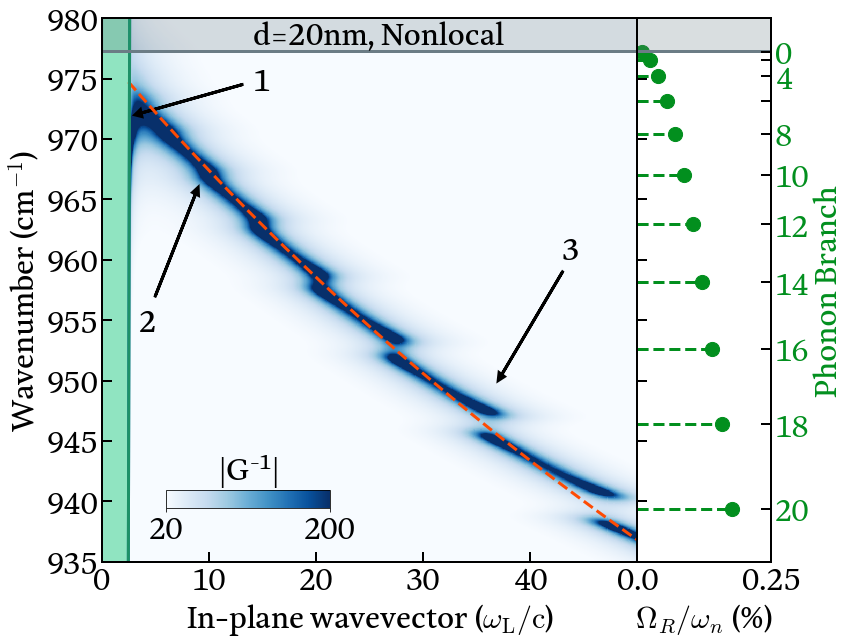}
	\caption{\label{fig:Fig3} Full dispersion relations from Eq.~\ref{eq:nldisp} for film thickness 20nm, right panel shows the predicted normalised Rabi frequencies $\Omega_R / \omega_n$. Labelled arrows refer to regions of the plot discussed in the text.}
\end{center}
\end{figure}\\

Having theoretically described the spectrum of the system, and shown the appearance of multiple strong-coupling features, we want now to derive from Eq.~\ref{eq:nldisp} a more transparent description of the nonlocality-mediated coupling.
The key step to pass from a holistic description to a modal one is to expand the hyperbolic function on the right of Eq.~\ref{eq:nldisp} as the Mittag-Leffler series
\begin{align}
	\frac{\tanh\left(\xi d / 2\right)}{\xi} &= \frac{d}{2} \sum_n \frac{8}{\left(2 n - 1\right)^2 \pi^2 + \xi^2 d^2},\nonumber \\
	&= \frac{d}{2} \sum_{n \in \mathrm{even}} \frac{8 \beta_{\mathrm{L}}^2 / d^2}{\omega \left(\omega + i \gamma\right) - \omega_{n}^2}, \label{eq:ml}
\end{align}
where we utilised Eq.~\ref{eq:LOFull} to eliminate $\xi$ and recognised in the denominator the discrete phonon frequencies Eq.~\ref{eq:phononfreq}. This is a sum over discrete phonon modes with frequency given by the even solutions to Eq.~\ref{eq:phononfreq}. Utilising Eq.~\ref{eq:ml} and the ENZ approximations discussed earlier we can transform the transcendental Eq.~\ref{eq:nldisp} into a factorised dispersion relation
\begin{equation}
    1 = \frac{\omega_{\mathrm{L}}^2 - \omega_{\mathrm{ENZ}, k_{\parallel}}^2}{\omega_{\mathrm{ENZ}, k_{\parallel}}^2 - \omega \left(\omega + i \gamma\right)} \sum_{n \in \mathrm{even}} \frac{8 \beta_{\mathrm{L}}^2 / d^2}{\omega_{n}^2 - \omega \left(\omega + i \gamma\right)} = 0, \label{eq:finaldisp}
\end{equation}
where $\omega_{\mathrm{ENZ}, k_{\parallel}}$ is the ENZ mode frequency from Eq.~\ref{eq:enzdisp} in the limit $\gamma \to 0$.
The polaritonic frequencies obtained from this equation are plotted as solid yellow lines in Fig.~\ref{fig:Fig2}b, e, for $\gamma \to 0$, demonstrating an excellent agreement between the polaritonic model and the full dispersion map. \\

In order to develop a modal, quantum description of the hybrid longitudinal-transverse polaritons which can be used to both directly link the system spectrum to the geometric and material device parameters, and to provide a way to calculate the rates of polariton generation and scattering,
we start by introducing second-quantized operators for the bare modes. We thus define 
$\hat{a}_{k_{\parallel}}^{\dag}$ as the creation operator for an ENZ mode with in-plane wavevector $k_{\parallel}$
and $\hat{b}_{n,k_{\parallel}}^{\dag}$ for the $n$th discrete longitudinal phonon mode of the layer. The coupling between resonances can be described by a generic, phenomenological Hamiltonian modeling resonant energy exchange between the transverse ENZ and the longitudinal phonons
\begin{multline}
	\mathcal{H}_{k_{\parallel}} = \hbar \omega_{\mathrm{ENZ}, k_{\parallel}} \hat{a}_{k_{\parallel}}^{\dag} \hat{a}_{k_{\parallel}} 
	+  \sum_{n \in \text{even}} \biggr[\hbar \omega_{n} \hat{b}_{n, k_{\parallel}}^{\dag} \hat{b}_{n, k_{\parallel}}\\
	 + \frac{\hbar \Theta_{n, k_{\parallel}}}{2} \left( \hat{a}_{-k_{\parallel}}^{\dag} +  \hat{a}_{k_{\parallel}} \right) \left( \hat{b}_{n, k_{\parallel}}^{\dag} +  \hat{b}_{n, -k_{\parallel}} \right)\biggr], \label{eq:ham}
\end{multline}
where $\Theta_{n, k_{\parallel}}$ is the $n$th phonon coupling frequency at common in-plane wavevector $k_{\parallel}$. The eigenvalue equation for this Hamiltonian can be written considering polariton operators of the form 
\begin{align}
\hat{d}_{k_{\parallel}} = \alpha_{k_{\parallel}} \hat{a}_{k_{\parallel}} + \zeta_{k_{\parallel}} \hat{a}_{-k_{\parallel}}^{\dag} + \sum_{n \in \mathrm{even}} \beta_{n, k_{\parallel}} \hat{b}_{n, k_{\parallel}} + \theta_{n, k_{\parallel}} \hat{b}_{n, -k_{\parallel}}^{\dag},
\end{align}
where Greek symbols are Hopfield coefficients \cite{Hopfield1958}. In order for $\hat{d}_{k_{\parallel}}$ to diagonalise the Hamiltonian $\mathcal{H}_{k_{\parallel}}$ it must satisfy the eigenequation 
\begin{equation}
    \left[\hat{d}_{k_{\parallel}} , \mathcal{H}_{k_{\parallel}} \right] = \hbar \omega \hat{d}_{k_{\parallel}} \label{eq:Poleom}
\end{equation}
where $\omega$ is the frequency of the coupled mode. By expanding the commutator in Eq.~\ref{eq:Poleom}, collecting terms proportional to each operator in Eq.~\ref{eq:ham} on each side and eliminating we recover the secular equation
\begin{align}
	1 = \frac{\omega_{\mathrm{ENZ}, k_{\parallel}}}{\omega_{\mathrm{ENZ}, k_{\parallel}}^2 - \omega^2} \sum_{n \in \mathrm{even}} \frac{\lvert \Theta_{n, k_\parallel} \rvert^2 \omega_{n}}{\omega_{n}^2 - \omega^2}, \label{eq:hamdisp}
\end{align}
whose roots provide the polariton eigenfrequencies of the system. Note that although we have ignored losses in the quantum model, leading to a disparity between Eq.~\ref{eq:hamdisp} and Eq.~\ref{eq:enzdisp}, they could be included by coupling Eq.~\ref{eq:ham} to a thermal bath \cite{Gubbin2016}.
By comparison with Eq.~\ref{eq:finaldisp} we can then directly derive one of the key results of this work, that is the analytical expression of the longitudinal-transverse coupling 
\begin{align}
	\lvert \Theta_{n, k_\parallel} \rvert^2 = \frac{\omega_{\mathrm{L}}^2 - \omega_{\mathrm{ENZ}, k_{\parallel}}^2}{\omega_{\mathrm{ENZ}, k_{\parallel}} \omega_{n}}  \frac{8 \beta_{\mathrm{L}}^2}{d^2}.
\end{align}
We can already see that the coupling increases linearly with $\beta_{\mathrm{L}}/d$, quantifying our physical intuition that the nonlocality-mediated coupling increases when phonon propagation cannot be neglected (large $\beta_{\mathrm{L}}$) and for smaller structures (small $d$).

To illustrate the physical content of Eq.~\ref{eq:finaldisp} we specialise onto the case where only the $n$th phonon branch is near-resonant with the ENZ mode and the others can be safely neglected, yielding polariton frequencies
\begin{align}
	\omega_{\pm, k_{\parallel}}^2 &= \frac{\omega_{\mathrm{ENZ}, k_{\parallel}}^2 + \omega_{n}^2}{2} \\
	& \pm \frac{\sqrt{\left(\omega_{\mathrm{ENZ}, k_{\parallel}}^2 - \omega_{n}^2\right)^2 + 4 \lvert \Theta_{n ,k_{\parallel}}\rvert^2 \omega_{\mathrm{ENZ}, k_{\parallel}} \omega_{n}}}{2}. \nonumber
\end{align}
The vacuum Rabi frequency, defined as half the resonant polariton splitting at the anticrossing $\omega_{\mathrm{ENZ}, k_{\parallel}} = \omega_{n}$, can thus be written as
\begin{align}
	\Omega_R &= \frac{\omega_{+, k_{\parallel}} - \omega_{-, k_{\parallel}}}{2}.
\end{align}
This quantity determines the coupling regime of the system. When $\Omega_R$, is smaller than the linewidths the system is in weak coupling, while when it is sizeably larger we can resolve the polaritonic splitting and the system is in the strong coupling regime. In the small coupling regime we can write the Rabi frequency, normalised by the anticrossing frequency
\begin{equation}
    \frac{\Omega_R}{\omega_{n}} = \frac{\sqrt{2}}{\xi_n d \left(\omega_{\mathrm{L}}^2 /\Delta \omega_n^2 - 1 \right)},\label{eq:redRabi}
\end{equation}
where the longitudinal mode frequency shift $\Delta \omega_n = \beta_{\mathrm{L}}^2\xi_n^2$ describes the departure in the $n$th phonon frequency from its zone-centre frequency. The result demonstrates two routes to increasing the Rabi frequency through tuning the phonon component, either through considering higher order (larger $n$) modes or exploiting materials with larger phonon velocity $\beta_{\mathrm{L}}$ to increase $\Delta \omega_n$.\\
 The $d^{-1}$ remaining in Eq.~\ref{eq:redRabi} formula arises from the ENZ and can be related to the out-of-plane electric field in the central layer as $\lvert \mathrm{E}_{z, k_{\parallel}} \rvert \propto d^{-1}$ following Campione \cite{Campione2015}. This implies that for a general structure the coupling strength can be enhanced by increasing the photonic field strength at the nonlocal interface. This could allow for larger photon-phonon coupling rates in nanophotonic structures where the photon field is confined in multiple dimensions through Purcell factor optimisation \cite{Caldwell2013, Gubbin2016}.\\

To explore the transition between the weak and the strong coupling regime, with the opening of the polaritonic gap,
we consider a wider structure, in order to have access to more phonon modes in the Reststrahlen region. In Fig.~\ref{fig:Fig3} we thus plot the dispersion of a 20nm 3C-SiC film, with three qualitatively different regions highlighted. In region 1 the ENZ approximation leading to Eq.~\ref{eq:ENZ} is not valid and the
coupled mode dispersion follows the bare ENZ dispersion, being initially parallel to the light-line. In region 2 larger $n$ causes an increase in ENZ linewidth where the ENZ dispersion crosses the localised phonon one. In this region localised phonons act as loss channels, extracting energy from the ENZ mode, but as $\Omega_R$ is smaller than the linewidth, the energy is lost before it can cycle back. In region 3 the polariton frequencies are sufficiently far from $\omega_{\mathrm{L}}$, strong coupling is established, and anti-crossings emerge in the spectra. The evolution of the normalised coupling $\Omega_R / \omega_{n}$ is shown in panel b, highlighting the expected increase for higher order modes.

\begin{figure*}[t!]
\begin{center}
	\includegraphics[width=0.8\textwidth]{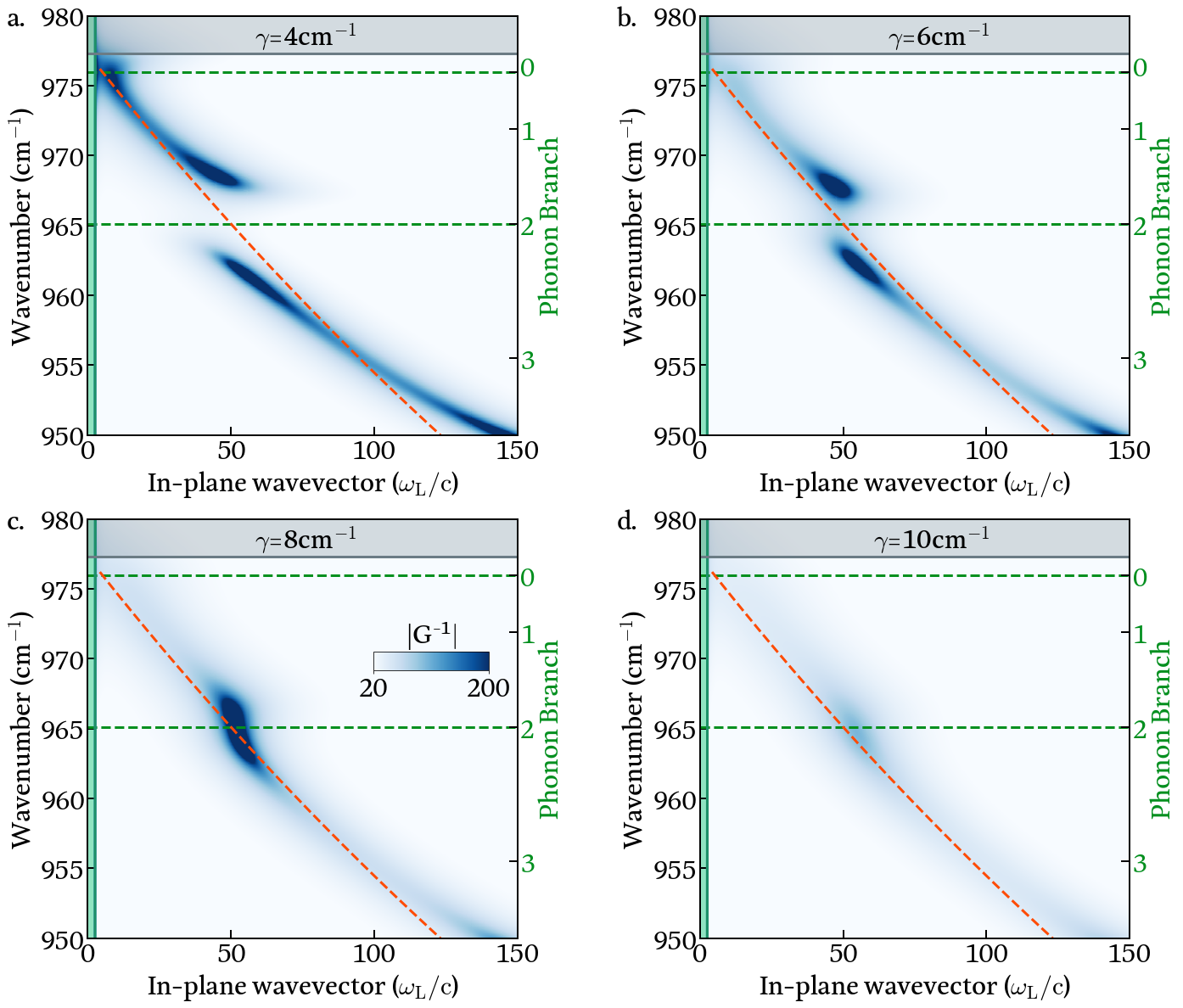}
	\caption{\label{fig:Fig4} Nonlocal dispersion relations for a 5nm 3C-SiC layer, analogous to Fig. 3d, for various values of the loss parameter $\gamma$. Panels show a. $\gamma = 4 \mathrm{cm}^{-1}$, b. $\gamma = 6 \mathrm{cm}^{-1}$, c. $\gamma = 8 \mathrm{cm}^{-1}$, d. $\gamma = 10 \mathrm{cm}^{-1}$.
	The red dashed line indicates the ENZ from Eq.~\ref{eq:enzdisp}. Increasing the losses the anticrossing is reduced and then disappears, as the system passes from the strong to the weak coupling regime.}
\end{center}
\end{figure*}

Having explored the weak-strong transition by varying the vacuum Rabi frequency while keeping the linewidths fixed, now we study the opposite case: increasing the losses while keeping fixed geometry and other material parameters determining the coupling strength. 
This is important not only in view of applications of our theory to other materials, but even for the case of 3C-SiC, which we chose for its preponderance as testbed for phonon-polariton physics \cite{Caldwell2013, Greffet2002, Hillenbrand2002, NeunerIII2009, Paarmann2016}.
We have in-fact used the 3C-SiC bulk value for $\gamma=2\mathrm{cm}^{-1}$ \cite{Moore1995},
but high-quality nanoscale 3C-SiC films are typically grown on silicon, and ultrathin films can be expected to have high defect densities, increasing material loss. We have moreover neglected that LO phonons are reported to have marginally larger losses than transverse ones \cite{Debernardi1999}.

To illustrate the effect of material loss on the nonlocal physics we study the $5$nm structure investigated earlier with increased damping rates. Results are shown in Fig.~\ref{fig:Fig4} for damping rates of $\gamma = 4, 6, 8, 10 \mathrm{cm}^{-1}$. Note that for this structure the vacuum Rabi frequency with the $n=2$ LO phonon branch is estimated through Eq.~\ref{eq:redRabi} as $\Omega_{\mathrm{R}} = 3.7 \mathrm{cm}^{-1}$. In panel a. where $\gamma \approx \Omega_{\mathrm{R}}$ the system response is very similar to that in Fig.~\ref{fig:Fig2}d where $\gamma = 2 \mathrm{cm}^{-1}$. The system remains in strong coupling. When $\gamma$ is increased to $6\mathrm{cm}^{-1}$ in panel b. the anti-crossing is reduced but still visible. In panels c. and d. where $\gamma$ is increased to $8\mathrm{cm}^{-1}$ and $10\mathrm{cm}^{-1}$ respectively, the anticrossing vanishes and the system passes into the weak coupling regime.
This is a combination of two physical effects, firstly the damping rate has increased with respect to the Rabi frequency and secondly the nonlocal figure of merit, which is the skin depth of the LO phonon $L = \mathrm{Im} \left\{\xi \right\}^{-1}$\cite{Gubbin2020} has decreased sufficiently such that the thin film no longer possesses a discrete phonon spectrum. When $\gamma = 4 \mathrm{cm}^{-1}$ we find $L = 1.3 d$, while when $\gamma = 10 \mathrm{cm}^{-1}$ we recover $L = 0.53 d$. At higher damping rates LO phonons can no longer transport energy between the two interfaces of the central film, instead they just act as an additional loss channel for the ENZ mode.\\
These effects can be mitigated utilising methods such as vapor-liquid-solid growth, which allow for the fabrication of high-quality sub-nanometer SiC films \cite{Sannodo2020}. Additionally alternative materials, particularly those which can be grown in high-quality thin films by atomic layer deposition on lattice matched substrates such as AlN or GaN can be utilised. Nonlocality in these anisotropic systems is expected to be physically similar to that presented in this work \cite{Gubbin2020c}. A second mitigation tactic is obvious through Eq.~\ref{eq:redRabi}, by decreasing the film thickness the coupling frequency is enhanced. This is a particularly promising route to explore nonlocal effects in systems grown by atomic layer deposition where high-quality films can be grown down to the true nanometer scale. For the parameters used in this work ($\beta_{\mathrm{L}} = 15.39 \times 10^5 \mathrm{cm \; s}^{-1}$, $\gamma = 2\mathrm{cm}^{-1}$) this yields an onset at $d \approx 4$nm for the $n=2$ phonon.\\

We have demonstrated that the full nonlocal Maxwell equations lead naturally to strong coupling between ENZ modes and localised phonon modes in a nonlocal dielectric nanolalyer. 
We developed an analytical second-quantized analytical approach quantitatively describing the system. Our theory provides a polaritonic description of the nanolayer physics which allows to naturally integrate polariton scattering and generation in a fully quantum picture. To this aim we exploit the Mittal-Leffler expansion as a link between the polynomial dispersion relations yielded by a coupled mode theory and those found using the full nonlocal Maxwell's equations, allowing us to condense the complexity of the nonlocality-mediated coupling in a single coupling frequency dependent on geometric and material parameters. We could thus directly calculate the minimal layer thickness, for a fixed material, leading to the appearance of strong coupling and LTP modes. Although the nonlocal effects studied in this paper occur far outside the light-line, they could be observed in a simple experiment by incorporating a metallic grating into the waveguide structure to fold the modes toward zone-centre \cite{Greffet2002, Gubbin2016}. Our technique, derived in a specific system, could be expanded to more general settings, providing a powerful modal picture to study nonlocal optical nanodevices.  

\section*{Acknowledgements}
The authors would like to thank Thomas G. Folland for fruitful discussions.

\section*{Funding}
\label{SecAck}
S.D.L. is supported by a Royal Society Research fellowship and the  the Philip Leverhulme prize. The authors acknowledge support from the Royal Society Grant No. RGF\textbackslash EA\textbackslash181001 and the Leverhumlme Grant No. RPG-2019-174.

\section*{Disclosures}
The authors declare no conflicts of interest.

\appendix

\bibliographystyle{naturemag}	
\bibliography{bibliography}

\end{document}